\documentclass[runningheads]{llncs}
\usepackage[T1]{fontenc}
\usepackage{color}

\usepackage{graphicx,verbatim}
\usepackage{booktabs}
\usepackage{multirow}
\usepackage{amsmath}
\usepackage{float}
\usepackage{bm}
\usepackage{amsfonts}
\usepackage{threeparttable}
\usepackage{hyperref}
\usepackage{xcolor}
\hypersetup{colorlinks=true, linkcolor=blue, citecolor=blue, urlcolor=blue}
\usepackage{marvosym}

\let\svthefootnote\thefootnote
\newcommand\freefootnote[1]{%
  \let\thefootnote\relax%
  \footnotetext{#1}%
  \let\thefootnote\svthefootnote%
}

\usepackage{perpage} 
\MakePerPage{footnote} 
\begin{document}

\title{Multiscale Causal Geometric Deep Learning for Modeling Brain Structure}

\author{Chengzhi Xia\inst{1}\textsuperscript{*} \and
    Jianwei Chen\inst{2}\textsuperscript{*} \and
    Yixuan Jiang\inst{1} \and
    Qi Yan\inst{3} \and
    Chao Li\inst{1,2,4}\textsuperscript{(\Letter)}}
\authorrunning{C. Xia et al.}
\institute{
    School of Science and Engineering, University of Dundee, Dundee, UK \and School of Medicine, University of Dundee, Dundee, UK \and College of Medicine and Biological Information Engineering, Northeastern University, Liaoning, China \and Department of Applied Maths and Theoretical Physics, University of Cambridge, Cambridge, UK \\
}


\maketitle
\begin{abstract}
Multimodal MRI offers complementary multi-scale information to characterize the brain structure. However, it remains challenging to effectively integrate multimodal MRI while achieving neuroscience interpretability. Here we propose to use Laplacian harmonics and spectral graph theory for multimodal alignment and multiscale integration. Based on the cortical mesh and connectome matrix that offer multi-scale representations, we devise Laplacian operators and spectral graph attentions to construct a shared latent space for model alignment.  Next, we employ a disentangled learning combined with Graph Variational Autoencoder architectures to separate scale-specific and shared features. Lastly, we design a mutual information-informed bilevel regularizer to separate causal and non-causal factors based on the disentangled features, achieving robust model performance with enhanced interpretability. Our model outperforms baselines and other state-of-the-art models. The ablation studies confirmed the effectiveness of the proposed modules. Our model promises to offer a robust and interpretable framework for multi-scale brain structure analysis.

\freefootnote{$*$ Equal contribution.}
\end{abstract}

\section{Introduction}
Multimodal MRI offers complementary information for characterizing brain structure. Structural MRI 
 (sMRI) reveals macrostructural features such as cortical volume, while diffusion MRI (dMRI) maps white matter microstructure. Integrating multimodal MRI provides a holistic view of brain structure across scales, enabling biomarkers for developmental and degenerative conditions, e.g., Alzheimer’s disease, which often exhibit multiscale structural abnormalities.

Challenges persist in integrating sMRI and dMRI for multiscale characterization of brain structure, due to their differences in resolution and dimensionality. Various multimodal fusion strategies have been proposed. For instance, Ning et al. \cite{ning2021relation} proposed a bidirectional mapping framework to align raw sMRI and dMRI in a shared latent space. Despite comprehensive features, this method may also introduce noise and redundancy. Similarly, Cai et al. \cite{cai2022graph} employed convolutional neural networks (CNN) 
extracting features from sMRI and dMRI to construct a graph neural network (GNN). Despite enhanced fusion, it may fail to accurately capture modality interaction and produce spurious connections. Further, CNN-extracted features often lack interpretability for biological understanding. 

A critical limitation of these methods is their reliance on raw MRI data \cite{wang2023multi,li2023g}, which is prone to noise and uncertainty, ultimately affecting model reliability. Importantly, they often lack neuroscience insights for modeling brain physiology and pathology. 
Geometric descriptors derived from sMRI, such as cloud points \cite{bello2020deep}, and triangular meshes \cite{smirnov2021hodgenet}, offer robust approaches for modelling complex macrostructural morphology. Similarly, connectome derived from dMRI effectively represents the brain microstructural connectivity \cite{wei2021quantifying}, capturing intricate connections between brain regions \cite{liu2024structural}. These geometric data with neuroscience priors promise to improve model robustness and interpretability. However, integrating the geometric representations of mesh and connectome remains challenging due to fundamental differences in scale and data structure. 

The Laplacian operator \cite{popov2022spectrum,pang2023geometric,anand2023hodge}, a fundamental mathematical tool, can capture high-frequency details and fine-grained structures for computer vision tasks, where its solutions, Laplacian harmonics \cite{ciaurri2017harmonic}, provide a mathematically robust and interpretable representation of object morphology. Neuroscience research has shown that Laplacian harmonics offer representations of brain function, with longer wavelengths corresponding to global brain activities and shorter wavelengths capturing localized activities. By integrating waves of different lengths, 
multiscale information of brain activity can be effectively captured. This capacity to span across scales presents a unique opportunity to bridge multiscale geometrics, enabling a comprehensive characterization of brain structure.

We propose a Laplacian MultiScale geometric learning (La-MuSe) framework to integrate mesh and connectome for characterizing brain structure. \textbf{Firstly}, to align the Laplacian harmonics derived from mesh and connectome, we introduce a latent space framework creating shared representations. 
Unlike static methods, our dynamic Laplacian decomposition leverages Spectral Graph Attention to iteratively refine Laplacian harmonics for optimal latent space alignment.
\textbf{Secondly}, to better integrate mesh and connectome while modelling their interactions, we employ a Cross-scale Disentangled Learning \cite{chai2021integrating} enhanced by a dual-branch Graph Variational Autoencoder (GraphVAE) bottleneck constraints to separate modality-specific and shared features.
\textbf{Thirdly},  inspired by MI \cite{zheng2024ci}, to enhance model reliability and interpretability, we propose a Bilevel  Mutual-information (MI)-regularized Causal Inference Module modelling cross-modality interactions and modality-specific causal graphs.
Our contributions are threefold:
\begin{itemize}
\item A latent space framework with spectral graph attention-based Laplacian decomposition to iteratively align cortical mesh and white matter connectome representations for multiscale modelling.
\item A GraphVAE bottleneck-constrained Cross-scale Disentangled Learning approach separates modality-specific and shared features, ensuring structured and biologically relevant multiscale integration. 
\item A Bilevel MI-regularized Causal Inference module modelling within-modality causal graphs and cross-modality interactions.
\end{itemize}
Our experiments demonstrate that our framework effectively integrate cortical meshes and connectome for downstream predictions, surpassing baseline and competing methods in age and sex predictions.

\begin{figure}[ht]
    \makebox[\textwidth][c]{
    \includegraphics[scale=0.7]{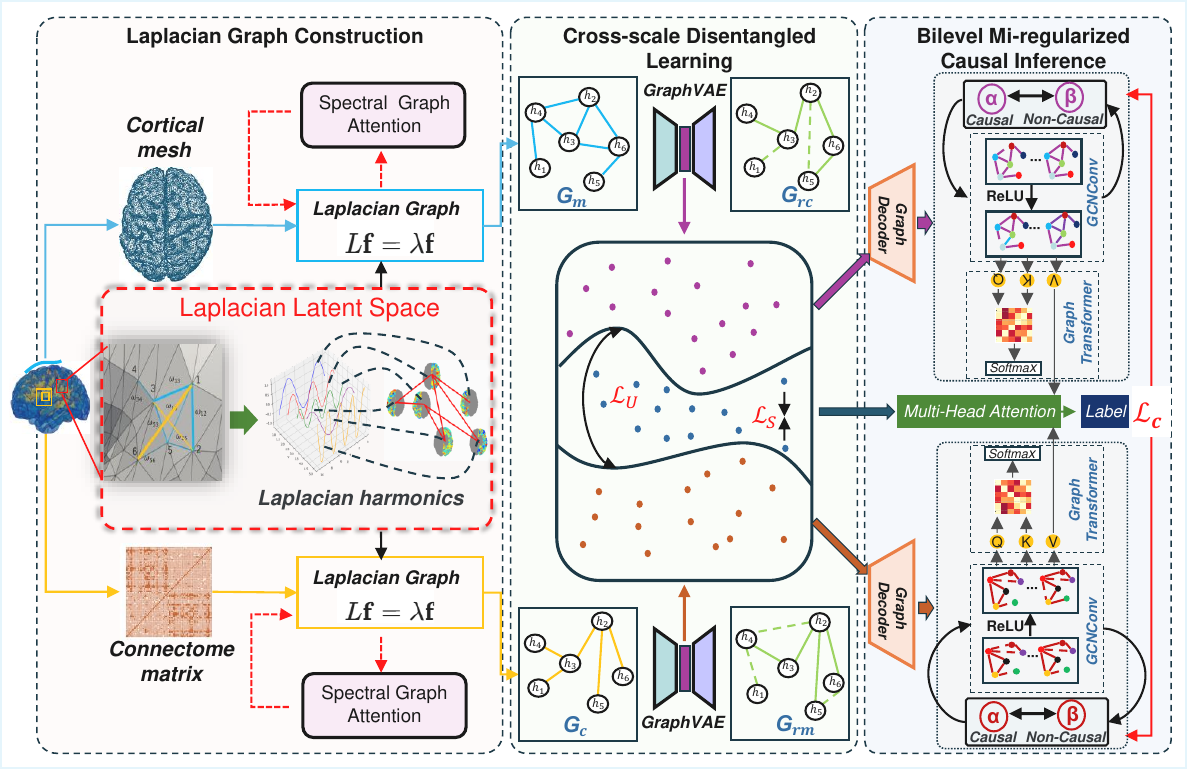}
    }
    \caption{The framework includes a Laplacian Latent Space with spectral graph attention  (left), a Cross-scale Disentangled Learning module (middle) and a Bilevel MI-regularized Causal Graph (right).}
    \label{fig1}
\end{figure}
\section{Methods}

The La-MuSe framework integrates a Laplacian Latent Space (2.1), a Cross-scale Disentangled Learning module (2.2), and a Bilevel MI-regularized Causal Inference module (2.3) to model multiscale brain structure. Firstly, we input the cortical mesh derived from T1-weighted MRI and connectome matrix derived from dMRI into the Laplacian latent space for modality alignment. The derived Laplacian matrices are projected onto their fused representation by Rayleigh quotient to establish initial graph structures, constructed via cosine similarity between dominant eigenvectors. 
These graphs undergo iterative refinement through Spectral Graph Attention which dynamically adjusts edge weights while preserving spectral properties. 
Next, those graphs of similar frequency domains in the same space are modelled by a Cross-scale Disentangled Learning module to derive scale-specific and shared geometric features. 
We introduce a dual-branch GraphVAE architecture to reconstruct one modality from the other to ensure structured shared information. 
Finally, we propose a Bilevel MI-regularized Causal Inference based on the disentangled modality-specific features to filter spurious graph connections. A regularizer models cross-modality interactions. The final features are fed into graph transformers for downstream predictions. 


\subsection{Laplacian Latent Space}
Given paired sMRI and dMRI, high-precision cortical meshes and connectome matrices are coregistered so that each vertex on the mesh corresponds to a node in the connectome. This allows to map the cortical geometrics with the structural connectivity. At each vertex, we compute the Laplacian harmonics as follows:\\
\begin{equation}
    \Delta f(\vec{v}_i) = \frac{1}{2A_i} \sum_{j \in \mathcal{K}(i)} (\cot \alpha_{ij} + \cot \beta_{ij})(f_j - f_i),
    \label{eq:placeholder}
\end{equation}
where $A_i$ denotes the Voronoi area at vertex $v_i$, and $\alpha_{ij}$, $\beta_{ij}$ are the angles opposite to the edge connecting $v_i$ and $v_j$, and $K$ is the number of mesh vertexes. The Laplacian matrix, $\mathbf{L}\in \mathbb{R}^{K\times K}$, is constructed based on the cotangent weights: $\mathbf{L}_{ij}$ is calculated by  $(\cot \alpha_{ij} + \cot \beta_{ij})$, while its diagonal entries $\mathbf{L}_{ii}$ is defined as the sum of the cotangent weights for each row.

Three Laplacian matrices are derived: $\mathbf{L}_m$ representing the cortical geometry, $\mathbf{L}_c$ representing the connectome matrix, and their fused representation $\mathbf{L}_{\text{fuse}}$, whose eigenvectors define the latent space.
FThe $\mathbf{L}_m$ and $\mathbf{L}_c$ are projected onto the latent space by the Rayleigh quotient \cite{li2015rayleigh} for 
eigenvector alignment as follows:  
\begin{equation}
    R(\mathbf{L}, \mathbf{x}) = \frac{\mathbf{x}^H\mathbf{L}\mathbf{x}}{\mathbf{x}^H\mathbf{x}},
    \label{eq:rayleigh_quotient}
\end{equation}  
where $\mathbf{x}$ represents harmonics of $\mathbf{L}_{\text{fuse}}$, $\bm{x}^H\in \mathbb{R}^{N\times F}$ is the conjugate transpose of $\bm{x}$, and $\mathbf{L}$ is $\mathbf{L}_c$ or $\mathbf{L}_m$. 

The alignment yields the corresponding eigenvalues and establishes initial graph structures, mesh graph $\mathbf{G}_m\in \mathbb{R}^{N\times F}$ and connectome graph $\mathbf{G}_c\in \mathbb{R}^{N\times F}$,
via cosine similarity between dominant eigenvectors, where $N$ is the number and $F$ is the length of Laplacian harmonics after initial projection simplification.

Finally, the eigenvalues multiplied by the eigenvector are used as nodes, while the cosine similarity is used as the initial edge to construct two Laplacian graphs. The initial graphs are iteratively refined through the Spectral Graph Attention network that preserves the spectral features:
\begin{equation}
   e_{ij} = \text{LeakyReLU}\left( \mathbf{a}^T \left[ \mathbf{W}\mathbf{x}_i \, \| \, \mathbf{W}\mathbf{x}_j \right] \right),
   \label{eq:gat}
\end{equation}
where $\mathbf{W} \in \mathbb{R}^{F \times F}$ is a learnable projection matrix, $\mathbf{a} \in \mathbb{R}^{2F}$ the attention vector. The edge weights are normalized as:
\begin{equation}
   \alpha_{ij} = \frac{\exp(e_{ij})}{\sum_{k \in \mathcal{N}(i)}\exp(e_{ik})}.
\end{equation}

\subsection{Cross-scale Disentangled Learning}
The shared eigenvectors align the mesh graph $\mathbf{G}_m$  and connectome graph $\mathbf{G}_c$ in the shared space, which, however, may not effectively leverage the complementary information from each modality. To mitigate this limitation, we devise a Cross-scale Disentangled Learning module to disentangle shared and unique features. 
To further facilitate effective modality fusion and alignment, we devise a dual-branch GraphVAE architecture~\cite{simonovsky2018graphvae}, which maps the shared information of meshes and connectomes through cross-reconstructing each modality from the latent representations of the other. This could ensure structured latent features.   

Specifically, shared features $\mathbf{Z}_m^{\text{sh}}, \mathbf{Z}_c^{\text{sh}} \in \mathbb{R}^{k \times d}$ capturing cross-modal correlations of $\mathbf{G}_m$ and $\mathbf{G}_c$, together with the modality-specific features, $\mathbf{Z}_c^{\text{uni}},\mathbf{Z}_m^{\text{uni}} \in \mathbb{R}^{k \times d}$, are derived through a standard encoder as follows:
\begin{equation}
    \begin{aligned}
        \mathbf{Z}_m^{\text{sh}},\mathbf{Z}_m^{\text{uni}} &= \text{Enc}_\phi(\mathbf{G}_m),
        \quad\mathbf{Z}_c^{\text{sh}}, \mathbf{Z}_c^{\text{uni}} = \text{Enc}_\phi(\mathbf{G}_c), \\
    \end{aligned}
\end{equation} 
where $\phi$ is the learnable parameters.
The reconstructed graphs, $\mathbf{G}_{rc}, \mathbf{G}_{rm} \in \mathbb{R}^{N \times N}$, enforcing feature orthogonality in shared space, are defined as:  
\begin{equation}
    \begin{aligned}
        \mathbf{G}_{rc} &= \text{Dec}_\theta(\mathbf{Z}_m^{\text{sh}}, \mathbf{Z}_c^{\text{uni}}), 
        \quad\mathbf{G}_{rm} = \text{Dec}_\theta(\mathbf{Z}_c^{\text{sh}}, \mathbf{Z}_m^{\text{uni}}),
    \end{aligned}
\end{equation} 
where $\theta$ is the learnable parameters. 
Finally, we design the disentangled loss as:
\begin{equation}
\label{eq:losscr}
\frac{\mathcal{L}_s}{\mathcal{L}_u} = \frac{\| \mathbf{Z}_m^{\text{sh}} - \mathbf{Z}_c^{\text{sh}} \|^2 + \| \mathbf{Z}_m^{\text{sh}} + \mathbf{Z}_c^{\text{sh}} - \mathbf{Z}_r^{\text{sh}}\|}{\| \mathbf{Z}_m^{\text{uni}} - \mathbf{Z}_c^{\text{uni}} \|^2},
\end{equation}
where $\mathbf{Z}_r^{\text{sh}} \in \mathbb{R}^{k \times d}$ is the union information of $\mathbf{G}_{rc}$ and $\mathbf{G}_{rm}$. This self-supervised module is jointly optimized with the causal module (2.3).



\subsection{Bilevel MI-regularized Causal Inference}
To effectively model causalities, we develop a Bilevel MI-regularized Causal Inference module, modelling causalities within and across modalities. 
Within each modality, the disentangled features $\mathbf{Z}^{\text{uni}}$ are decomposed into causal ($\boldsymbol{\alpha}$) and non-causal ($\boldsymbol{\beta}$) features:  
\begin{equation}
    \boldsymbol{\alpha} = \sigma(\mathbf{W}_g\mathbf{Z}^{\text{uni}}) \odot \mathbf{Z}^{\text{uni}},\quad\boldsymbol{\beta} = \mathbf{Z}^{\text{uni}} - \boldsymbol{\alpha},
\end{equation}  
where $\mathbf{W}_g\in \mathbb{R}^{d \times d}$ is a trainable projection matrix.

The graph transformer architectures are embedded with the causal inference module through learnable gating,taking causal features $\boldsymbol{\alpha}_m, \boldsymbol{\alpha}_c$ as inputs through multi-head graph attention with spectral positional encoding:  
\begin{equation}
\begin{aligned}
\mathbf{Q} = \boldsymbol{\alpha} \mathbf{W}_Q,\quad \mathbf{K} &= \boldsymbol{\alpha} \mathbf{W}_K,\quad \mathbf{V} = \boldsymbol{\alpha} \mathbf{W}_V, \\
\text{Attn}(\mathbf{Q}, \mathbf{K}, \mathbf{V}) 
    &= \text{Softmax}\left(\frac{\mathbf{Q}\mathbf{K}^T}{\sqrt{d_k}} \oplus \mathbf{B}_{\phi}\right)\mathbf{V},
\end{aligned}
\end{equation}
where $\mathbf{B}_{\phi} \in \mathbb{R}^{k \times k}$ contains spatial distances between cortical regions.
 
A regularizer term is further designed to model interactions between two causal graphs. Therefore, the bilevel causal loss is expressed as follows:

{\small
\begin{equation}
\mathcal{L}_{c} = 
\underbrace{-I(\boldsymbol{\alpha}_m; Y) - I(\boldsymbol{\alpha}_c; Y)}_{\text{within-modality causal}} 
+ \underbrace{\eta I(\boldsymbol{\alpha}_m;\boldsymbol{\beta}_m) + \eta I(\boldsymbol{\alpha}_c; \boldsymbol{\beta}_c)}_{\text{within-modality non-causal}}
+ \underbrace{\gamma I(\boldsymbol{\alpha}_m;\boldsymbol{\alpha}_c)}_{\text{cross-modality regularizer}},
\end{equation}
}
where $I(\cdot;\cdot)$ denotes MI , and $\eta, \gamma$ are regularization coefficients. The within-modality causal terms ensure that the model maximizes the causal relationship between the graph and predictions, while the within-modality noncausal terms minimize spurious connections. Together, they model the causality within each modality, while the cross-modality regularizer models their interactions.

Finally, the outputs of both modalities are integrated via: 
\begin{equation}
\mathbf{Y}_{\text{pred}} = \text{LayerNorm}\left(\text{Attn}(\mathbf{H}_m \mathbf{W}_m, \mathbf{H}_c \mathbf{W}_c) + \mathbf{Z}_r^{\text{sh}}\right),
\end{equation} 
where $\mathbf{W}_m, \mathbf{W}_c \in \mathbb{R}^{d \times d}$ are modality-specific projection matrices. The predicted Y is iteratively predicted to optimise the causal inference.


\section{Experiments and Results}
\subsection{Datasets and Preprocessing}

We included the first 115 subjects from the HCP datasets. Firstly, all T1 and dMRI were registered to the MNI152 space \cite{fonov2009unbiased}. The T1 images were processed using Freesurfer's \cite{FISCHL2012774} 'recon-all' pipeline before extracting cortical meshes from the cortical surface files generated by Freesurfer. The dMRI were processed by MRTRIX3 \cite{TOURNIER2019116137} with eddy current and geometric distortion correction. Then, we used 5-tissue-type (5tt) segmentation with Freesurfer outputs to generate a gray matter-white matter interface. Finally, we performed fiber tracking using the iFOD2 algorithm with MRTRIX3's ACT function. 

\subsection{Implementation and Evaluation}

The proposed method is implemented on a single GPU (NVIDIA GeForce RTX 3070Ti) with Pytorch and PyTorch Geometric. We trained our model using the Adam optimizer with epochs of 300, a learning rate of 0.001 and a mini-batch of 8. We used age prediction and gender prediction tasks to evaluate our model. For classification, we used accuracy (ACC) and F1 score, while regression performance was evaluated using mean absolute error (MAE) and root mean square error (RMSE).  Each dataset was split for 5-fold cross-validation, and results are reported as the mean and standard deviation across folds.  MSE and BCEWithLogits were chosen as the loss function of the graph transformer.

\begin{figure}[ht]
    \makebox[\textwidth][c]{
    \includegraphics[scale=0.45]{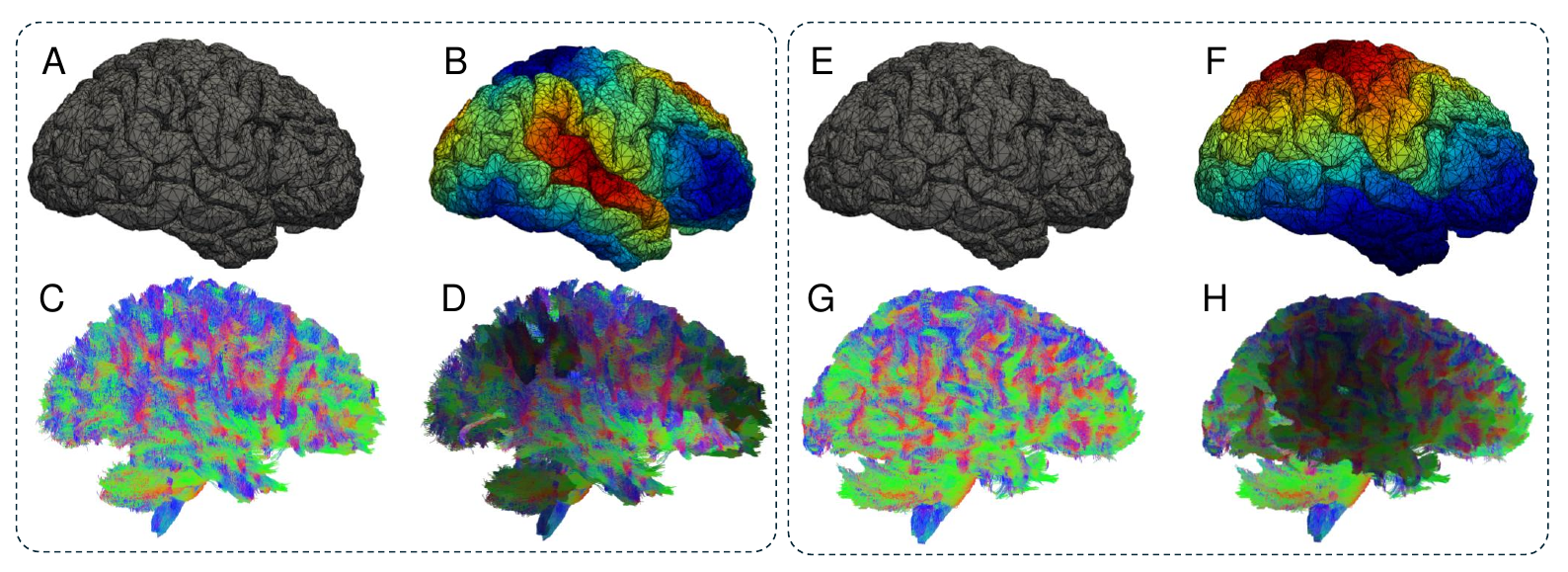}
    }
    \caption{Example heatmaps of mesh and connectome harmonics in age prediction on the HCP dataset. On both subjects,  A \& E show the original cortical meshes while C \& G visualise the structural connectomes. From the derived harmonic heatmaps of mesh (B \& F) and connectome (D\&H), younger (left) and older (right) subjects demonstrate distinct patterns in both mesh and connectome harmonics.}
    \label{fig1}
\end{figure}
\subsection{Model Comparisons}
We compared our model with baselines and SOTA models on raw sMRI/dMRI and mesh/connectome. For both raw data and geometric data,  performances are compared in both single-modal and multimodal settings. The results show that our model outperforms all the baseline methods and most SOTA methods on both age prediction and gender classification. For age estimation, our model achieves an MAE of 2.36 and an RMSE of 2.90, outperforming other models by at least 0.26 and 0.36 respectively. In gender classification, the full model records an accuracy of 79\% and an F1 score of 0.69, showing the highest accuracy among all models, outperforming other multimodal models by an average of 4.14\% and unimodal models by 9\%. Although other SOTA method report competitive F1 metrics in gender prediction, our model generally shows consistently higher performance.
\begin{table}[ht]
    \centering
    \caption{Comparisons with baselines and state-of-the-art models}
    \label{tab:results}
    \begin{tabular}{lcccccc}
        \toprule
         \multirow{2}{*}{\textbf{Method}} & \multirow{2}{*}{\textbf{Modality}} & \multicolumn{2}{c}{\textbf{Age}} & \multicolumn{2}{c}{\textbf{Gender}}\\ 
\cline{3-6} &  &  \textbf{MAE~$\downarrow$} & \textbf{RMSE~$\downarrow$} & \textbf{ACC~$\uparrow$} & \textbf{F1~$\uparrow$} \\
        \midrule
        \textbf{Ours}         & Mesh+Connectome      & \textbf{2.36(0.21)} & \textbf{2.90(0.49)} & \textbf{0.79(0.17)}  & 0.69(0.29) \\
        Cai, et al\cite{cai2022graph}     & Raw T1+DTI    & 2.78(0.11) &3.42(0.36)  & 0.77(0.17)  & 0.75(0.15)\\
        Ning, et al\cite{ning2021relation}   & Raw T1+DTI    & 3.01(0.43) & 3.56(0.41) & 0.76(0.11))  & 0.68(0.23) \\
        Cai, et al\cite{cai2022graph}     & Mesh+Connectome       & 2.63(0.22) &  3.41(0.29) & 0.78( 0.26) &  \textbf{0.76(0.33)}\\
        Yang, et al\cite{yang2023mapping}    & Mesh+Connectome      &2.62(0.63) &3.26(0.80) &0.74(0.09)  &0.68(0.25) \\
        \midrule
        \multirow{3}{*}{Transformer\cite{jun2023medical}}   & Raw T1+DTI      & 2.90(0.62) & 3.79(0.46) & 0.73(0.21)  & 0.71(0.20) \\
                      & Raw T1        & 3.04(0.55) & 3.85(0.42) & 0.65(0.19)  & 0.55(0.31) \\
                      & Raw DTI       & 2.91(0.61) & 3.99(0.45) & 0.77(0.13)  & 0.66(0.33) \\
        \midrule
        \multirow{3}{*}{CNN}   & Raw T1+DTI      & 2.82(0.94)  & 3.50(0.76) &0.73(0.17)  &0.71(0.15) \\
                      & Raw T1\cite{dartora2024deep}        & 2.98(0.22) & 3.73(0.17) & 0.69(0.13)  & 0.55(0.30) \\
                      & Raw DTI\cite{wang20233dcnn}       & 4.41(0.85) & 5.08(0.94) & 0.64(0.17)  & 0.55(0.33) \\
        \midrule
        \multirow{3}{*}{GCN}   & Mesh+Connectome      &3.05(0.48) &3.81(0.60) &0.73(0.14) &0.66(0.20)         \\
                      & Mesh         &3.29(0.56) &4.01(0.59) &0.70(0.12) &0.61(0.17)  \\
                      & Connectome        &3.42(0.56)  &4.19(0.60)  &0.75(0.13) &0.74(0.13)           \\
        \bottomrule
    \end{tabular}
\end{table}

\subsection{Ablation Studies}
Ablation studies (removing the proposed modules, laplacian graph construction, disentangled modeling, or causal network, one by one)  demonstrate the effectiveness of each module. As shown in Table 2, the complete model outperforms all its ablation variants by at least 8\% in ACC and by at least 0.19 in MAE. This demonstrates that all the proposed modules are effective. Of note, the model ablating the Cross-scale Disentanglement module performed the worst, with comparable performance with the unimodal baseline. This result indicates that the disentanglement module could guide the model to effectively leverage the harmonics of different scales.

\begin{table}[ht]
    \centering
    \begin{threeparttable}
    \caption{Results of ablation atudies}
    \label{tab:ablation}
    \begin{tabular}{lccccc}
        \toprule
        \multirow{2}{*}{\textbf{Methods}} & \multicolumn{2}{c}{\textbf{Age}} & \multicolumn{2}{c}{\textbf{Gender}}\\ 
\cline{2-5} 
         & \textbf{MAE~$\downarrow$} & \textbf{RMSE~$\downarrow$} & \textbf{ACC~$\uparrow$} & \textbf{F1~$\uparrow$} \\
        \midrule
        \textbf{Ours} & \textbf{2.36(0.21)} & \textbf{2.90(0.49)} & \textbf{0.79(0.17)} & \textbf{0.69(0.29)} \\
        \midrule
        \textit{w/o } Laplacian graph construction & 2.92(0.29) & 3.54(0.72) & 0.65(0.20) & 0.58(0.29) \\
        \textit{w/o } Cross-scale disentanglement & 3.06(0.24) & 3.64(0.53) & 0.62(0.19) & 0.53(0.23) \\
        \textit{w/o } Bilevel causal graph & 2.55(0.37) & 3.19(0.30) & 0.71(0.16) & 0.67(0.23) \\
        \bottomrule
    \end{tabular}
    \end{threeparttable}
\end{table}

\section{Conclusion}
In this study, we introduce a multiscale causal geometric deep learning framework that leverages Laplacian harmonics to effectively integrate cortical meshes and white matter connectome for modelling brain structure. Our model incorporates three key components: laplacian graph construction to capture frequency-domain features and align them in the shared latent space, a cross-scale disentangled learning module to separate shared and scale-specific information, and a bilevel MI-regularized causal inference module to model cross-scale dependencies while enhancing interpretability. Extensive experiments on age prediction and gender classification tasks demonstrate that our model consistently outperforms baseline and SOTA models. Our model offers an approach to stably characterize brain structure at different scales. Future work will further develop the model for characterizing brain physiology and pathology.




%
%
%
\bibliographystyle{splncs04}
\bibliography{reference}
\end{document}